\documentclass[letterpaper, 10 pt, conference]{ieeeconf}  

\IEEEoverridecommandlockouts                              
\usepackage[table,xcdraw]{xcolor}
\usepackage[english]{babel}
\usepackage{float}
\usepackage{subfig}
\usepackage{color}
\usepackage{bbm}
\usepackage{amsmath, bm}
\usepackage{multirow}
\usepackage{epsfig}
\usepackage{amsmath}
\usepackage{mathtools}
\usepackage{algorithm}
\usepackage{algorithmic}
\usepackage{amssymb}
\usepackage{verbatim}
\usepackage{lipsum}
\usepackage{graphicx}
\usepackage{cuted}
\usepackage{cite}
\usepackage{ulem}
\usepackage{url}

\newcommand{\tb}{\textcolor{black}}
\newcommand{\trd}{\textcolor{black}}
\newcommand{\trdd}{\textcolor{black}}
\newcommand{\ic}{\textcolor{black}}
\newcommand{\icc}{\textcolor{black}}

\newcommand{\sd}{\textcolor{black}}
\newcommand{\sdd}{\textcolor{black}}
\newcommand{\nic}{\textcolor{black}}
\newcommand{\fdic}{\textcolor{black}}

\DeclareMathOperator{\E}{\mathbb{E}} 

\title{\LARGE \bf AX-DBN: An Approximate Computing Framework for the Design of Low-Power Discriminative Deep Belief Networks}

\author{Ian Colbert, Ken Kreutz-Delgado and Srinjoy Das$^{1}$
\thanks{*Partially supported by NSF \& UCSD Calit2 Pattern Recognition Lab.}
\thanks{$^{1}$Communicating Author; email: s2das@ucsd.edu. All authors are affiliated with UC San Diego Pattern Recognition Lab.}
}

\begin{document}
\maketitle
\thispagestyle{empty}
\pagestyle{empty}

\vspace*{-0.65cm}
\begin{abstract}

The power budget for embedded hardware implementations of Deep Learning algorithms can be extremely tight. To address implementation challenges in such domains, new design paradigms, like Approximate Computing, have drawn significant attention. Approximate Computing exploits the innate error-resilience of Deep Learning algorithms, a property that makes them amenable for deployment on low-power computing platforms. This paper describes an Approximate Computing design methodology, AX-DBN, for an architecture belonging to the class of stochastic Deep Learning \trdd{algorithms} known as Deep Belief Networks (DBNs). Specifically, we consider procedures for efficiently implementing the Discriminative Deep Belief Network (DDBN), a stochastic neural network which is used for classification tasks, extending Approximation Computing from the analysis of deterministic to stochastic neural networks.
For the purpose of optimizing the DDBN for hardware implementations, we explore the use of: (a) Limited precision of neurons and functional approximations of activation functions; (b) Criticality analysis to identify the nodes in the network which can operate at reduced precision while allowing the network to maintain target accuracy levels; and (c) A greedy search methodology with incremental retraining to determine the optimal reduction in precision for all neurons to maximize power savings. 
Using the AX-DBN methodology proposed in this paper, we present experimental results across several network architectures that show significant power savings under a user-specified accuracy loss constraint \trdd{with respect to ideal full precision implementations}.

\end{abstract}

\section{Introduction}

In recent years, there has been a significant increase in the use of Deep Learning algorithms for a variety of cognitive computing applications such as image recognition, text retrieval and pattern completion~\cite{hinton2006reducing, salakhutdinov2007restricted, larochelle2012learning}. 
Enabling such algorithms to operate on low-power, real-time platforms such as
mobile phones and Internet of Things (IoT) devices is an area of critical interest. 
On such platforms, a training procedure is typically performed on a cloud server. This training involves optimizing the parameters of a cloud-based neural network using data presented to the device and uploaded to the cloud. Thereafter, the cloud performs classification when requested by the device, and subsequently the device communicates with the cloud each time an inference task is requested. This purely cloud-based approach to performing inference has a number of drawbacks including high power consumption, latency, security, and reliance on stable and fast Internet connectivity, and is therefore not suitable for use on ultra low-power devices with battery life constraints. Implementing inference locally on embedded hardware is perceived to be a desirable solution to this problem and is the focus of current research~\cite{venkataramani2014axnn,zhang2015approxann,das2015gibbs,chen2017eyeriss}. 

\begin{figure}[!t]
\centering
\includegraphics[width=12pc]{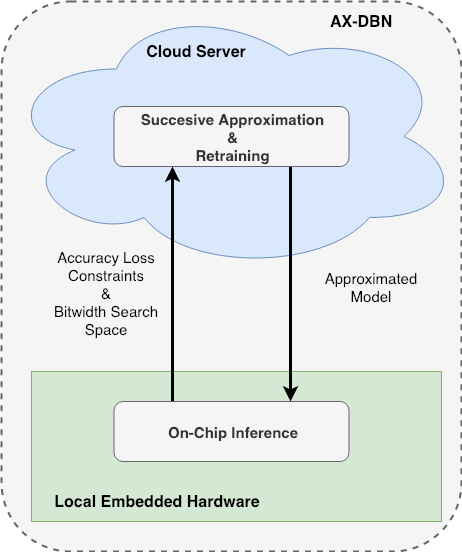}
\caption{\footnotesize \textsf{\textbf{Information Flow of the AX-DBN Methodology.} The macro-level design of our framework leaves all training, approximation, and retraining to the cloud server. The hardware client sends a request to the cloud server giving it an accuracy loss constraint and a fixed bitwidth search space. \tb{The resulting approximated model is then used for energy efficient inference on embedded hardware.}}}
\label{fig: macro info flow graph}
\end{figure}

We propose a shared approach where the cloud performs neural network training while a low-power implementation of the neural network on a local device is \sd{used for performing inference}, as shown in Fig.~\ref{fig: macro info flow graph}. Neural networks are inherently error-resilient~\cite{bishop1995neural} and, therefore, high precision of arithmetic representations and operations is not necessary to generate sufficiently accurate performance of such algorithms. 
This property is exploited by Approximate Computing techniques based on the use of limited precision representations and algorithm-level approximations to achieve energy efficient implementations with negligible performance loss~\cite{venkataramani2014axnn,zhang2015approxann}. 
In this paper, we \trd{propose AX-DBN}, an Approximate Computing methodology for a class of stochastic neural networks known as Discriminative Deep Belief Networks (DDBNs).
\fdic{These are multi-layered extensions of the Discriminative Restricted Boltzmann Machine (DRBM)~\cite{larochelle2012learning}}, which can be used for classification in both supervised and semi-supervised settings.
\trd{Our proposed AX-DBN framework extends the analysis of deterministic networks~\cite{venkataramani2014axnn, zhang2015approxann} to the domain of stochastic neural networks. AX-DBN involves training and approximating on the cloud and performing classification on embedded hardware, as depicted in Fig.~\ref{fig: macro info flow graph}.}
In \trd{this} approach, efficiency arises from exploiting arithmetical and functional approximations subject to a user-specified accuracy loss constraint relative to an ideal \trd{full precision} implementation.


\noindent{\bf Related Prior Work.}$~~$Previous work for neural network implementations in hardware using Approximate Computing have focused on feedforward deterministic networks solving classification tasks~\cite{zhang2015approxann,venkataramani2014axnn}. Venkataramani et al.~\cite{venkataramani2014axnn} propose a design-space exploration framework, AxNN, that systematically applies approximation techniques to deterministic neural networks. 
\icc{Using ``resilience characterization," their approach identifies and applies limited precision approximation on neurons whose impact will be the lowest on overall network performance to enable power optimized classification.}
Zhang et al.~\cite{zhang2015approxann} follows a similar approach where the formulation of their neuron criticality analysis is applied to identify neurons where approximate multiplier circuits are instantiated to reduce power during classification in feedforward neural networks. \sd{Their Approximate Computing framework ApproxANN, along with their power model, addresses energy savings for memory accesses and computation workloads based on network structures and hardware characteristics.} 

\noindent{\bf Contributions of this Paper.}$~~$\fdic{Extending the work of~\cite{venkataramani2014axnn} and~\cite{zhang2015approxann} into the stochastic domain, we develop a methodology allowing for the design of an energy efficient implementation of a stochastic Discriminative DBN (DDBN) to balance the trade-off between performance and power.\footnote{Although the stochastic Deep Belief Network can perform both classification and generation, in this paper we analyze its classification properties, leaving an analysis of its generation properties for future work.}}
Power efficiency arises from selectively exploiting arithmetical and functional approximations subject to a user-specified accuracy loss constraint. To accomplish this, two different levels of approximation are considered in the hardware implementation of a Discriminative DBN: (1) limited precision representation of hidden neurons; and (2) \icc{functional approximation of activation functions.} 
The main contributions of our work are as follows:
\begin{itemize}
\item The use of criticality analysis to rank order neurons based on their contribution to DDBN network performance.
This work carries criticality analysis to the domain of stochastic Deep Learning algorithms implemented on finite precision digital hardware.
\nic{Criticality metric driven approximation using Cross Entropy is compared against random ordering using Monte Carlo simulations to gauge their effectiveness in limited precision network approximation with variable bitwidths for individual neurons.}
\item The use of a greedy retraining procedure to optimize neuron bitwidths under given accuracy loss constraints with respect to ideal full precision implementations.
\item \sd{The use of a generalized power model for both computation workloads and memory accesses based on fixed point representations of \fdic{individual neurons}, number of samples and hardware characteristics.}
\end{itemize}

\noindent{\bf Outline of the Paper.}$~~$In Section II we review Discriminative RBMs (DRBMs) and describe how they are stacked to form Discriminative Deep Belief Networks (DDBNs). We also present the \sd{network} structures that are implemented in this paper. 
Section III describes how limited precision and activation function approximation will be performed for the purpose of implementing \sd{such networks} on hardware. 
In Section IV we present the mathematics for gradient-based neuron criticality analysis in stochastic neural networks such as \sd{DRBMs and DDBNs}.
In Section V we outline our AX-DBN hardware design methodology.
Section VI outlines the power model used for this work.
Section VII describes the accuracy and power results for different DRBM and DDBN architectures using our design methodology and resulting conclusions.
This work is a significant extension and improvement from our preliminary results reported in~\cite{xu2017approxdbn}.

\section{DISCRIMINATIVE DEEP BELIEF NETWORKS AND NETWORK ARCHITECTURE}
\label{sec: dbn theory}

\fdic{Here, we review the mathematical basics of the DRBM, the building block of DDBNs, and present the DRBM and DDBN configurations used for this work.}\\


\begin{figure}[t]
\centerline{\includegraphics[width=13pc]{./FIGS/RBM-Architecture.png}}
\caption{\footnotesize \textsf{\textbf{Discriminative RBM Architecture}: The visible neurons $\bm{v}$, made up of input neurons $\bm {x}$ and classification neurons $\bm {c}$, are connected to the hidden neurons $\bm {h}$ by weights $\mathbf{W}$ and $\mathbf{W}_c$, respectively.}}
\label{disRBM}
\end{figure}

\noindent{\it A. Discriminative Restricted Boltzmann Machines.}
A Restricted Boltzmann Machine (RBM) is a generative stochastic artificial neural network that can learn a probability distribution over its set of inputs. The RBM is a bipartite graph that consists of two layers - one visible layer $\bm v$ where the states of the units in this layer can be observed, and one hidden layer $\bm h$.
\fdic{The probability distribution implemented by the RBM is given by the Boltzmann distribution defined by:}

{\small{\begin{equation}
\label{eq:RBMp_xhc}
P(\bm v,\bm h) = \frac{e^{-E(\bm v,\bm{h})}}{\sum\limits_{\bm v,\bm{h}} e^{-E(\bm v,\bm{h})}} 
\end{equation}}}

\vspace{-0.3cm}
\noindent{\fdic{where the energy function $E(\bm{v},\bm{h})$ is given below:}}

{\small
\begin{equation}
\label{eq:RBMenergyfunction}
E(\bm v,\bm h) = - \bm v^T \bm b^v - \bm h^T \bm b - \bm v^T \bm W \bm h
\end{equation}}

\vspace{-0.5cm}
{Here, $\bm b^v$ and $\bm b$ are biases for the visible and hidden layers, respectively, and $\bm W$ is the weight matrix between them~\cite{hinton2010practical}. From Eq.~\ref{eq:RBMp_xhc}, one can derive the conditional distributions:}

\vspace{-0.3cm}
{\small{\begin{align}
\label{eq:RBMPh}
&P(h_j|\bm v)=\sigma(b_j + \sum\limits_i v_i W_{ij})\\
\label{eq:RBMPv}
&P(v_i|\bm h)=\sigma(b^v_i + \sum\limits_j h_j W_{ij})
\end{align}}}

\vspace{-0.3cm}
\noindent{where $\sigma (x)$ is the logistic sigmoid function $1/(1+\exp(-x))$. The training method we use is described in~\cite{hinton2010practical}, where the Contrastive Divergence (CD) method is \tb{applied} to provide weight \trd{and bias} updates.}
To use RBMs as classifiers, Larochelle et al.~\cite{larochelle2012learning} proposed the use of Discriminative RBMs (DRBMs), which jointly concatenate the input $\bm x$ and its associated ``one-hot'' target class $\bm c$ to form the single visible layer $\bm v$, as illustrated in Fig.~\ref{disRBM}. The energy function defined by~\cite{larochelle2012learning} is given below:

\vspace{-0.25cm}
{\small \begin{equation}
\label{eq:DRBMenergyfunction}
E(\bm{x},\bm h, \bm c) = - \bm x^T \bm b^x - \bm h^T \bm b - \bm c^T \bm b^c - \bm x^T \bm W \bm h - \bm c^T\bm  W^c \bm h
\end{equation}}
\hspace*{-1.2mm}
\trdd{Here,} $\bm b^x$, $\bm b$ and $\bm b^c$ are biases \trdd{of the visible, hidden, and classification layers, respectively}. $\bm W$, $\bm W^c$ are the weight matrices between layers as shown in Fig.~\ref{disRBM}. The inputs $\bm x$ to the DRBM can be either binary or continuous. \fdic{In the scope of this work, we consider binary inputs $\bm x$ as they can be used for efficient implementations on finite precision hardware with low power and area.}

\trd{Similar to the RBM,} the following conditional distributions \trd{can be derived for \trdd{a} DRBM \trdd{with $V$ visible, $H$ hidden and $C$ class units}}~\cite{larochelle2012learning}:

\vspace{-0.5cm}
{\footnotesize \begin{align}
\label{eq:DRBMPh}
&P(h_j|\bm v)=P(h_j|\bm x, \bm c)=\sigma(b_j + \sum_{i=1}^V x_i W_{ij} + \sum_{i=1}^C c_i W^c_{ij})\\
\label{eq:DRBMPx}
&P(x_i|\bm h)=\sigma(b^x_i + \sum_{j=1}^{H} h_j W_{ij})\\
\label{eq:DRBMPc}
&P(c_i|\bm h)=\text{softmax}(b^c_i + \sum_{j=1}^{H} h_j W^c_{ij})
\end{align}}

\vspace*{-0.3cm}
\noindent where $\sigma (x)$ is the logistic sigmoid function $1/(1+\exp(-x))$ and softmax$(x_i)=\exp(x_i)/\sum_{j=1}^k \exp(x_j)$ \trdd{where $i \in \{1, \cdots, k \}$}. Similar to the RBM, the DRBM \trd{can also be} trained using Contrastive Divergence 
\trd{to estimate the network weight and bias values~\cite{larochelle2012learning}}. 
\icc{Using the notation developed by Larochelle et al.~\cite{larochelle2012learning}, the DRBM conditional probability of a class label $c_i$ given input $\bm x$ is}

{\small
\begin{equation}
\label{eq:pc}
P(c_i|\bm x)=\frac{e^{-F(\bm x, c_i)}}{\sum_{j=1}^{C} e^{-F(\bm x, c_j)}}
\end{equation}
}

\vspace{-0.2cm}
\noindent{where $F(\bm x, c_i)$ denotes the Free Energy of the DRBM given input $\bm x$ and class label \trdd{$c_i$}, \trd{as defined below:} }
{\small
\begin{equation}
\label{eq:FE}
F(\bm x, c_i) = -b^c_{c_i} - \sum_{j=1}^H \log\big(1 + \exp(b_j + W^c_{jc_{i}} + \sum_{i=1}^V W_{ji}x_{i})\big)
\end{equation}
}

\vspace{-0.35cm}
A trained DRBM \trd{can be used} to perform classification using two \trd{equivalent} methods~\cite{hinton2006fast}: 

\begin{itemize}
\item {\it Free Energy:} From Eq.~\ref{eq:pc}, it can be seen that for a given \trd{visible vector} \trd{$\bm x$} the class $\bm c$ that has the highest probability of activation corresponds to the minimum value of $F(\bm x, c_i)$. Therefore, in this method, the Free Energy for a given \trd{$\bm x$} is calculated for all labels $c_i$. The classification result is the class $c_i$ that corresponds to the minimum Free Energy.


\item {\it Gibbs Sampling:} The binary activation state of each class can also be found by repeatedly sampling all class neurons of the DRBM for a given visible vector. The class with the highest activation frequency given a sufficient number of sampling iterations is the classification result.\\


\end{itemize}

\vspace{-0.3cm}
\trdd{The Gibbs Sampling method, with suitable approximations for the sigmoid function, is more amenable to realization on digital hardware than Free Energy \fdic{due to the highly nonlinear dependence of Free Energy on its input values}.} 
\trdd{Therefore} in this work, we use Gibbs Sampling for inference in our hardware implementations of DRBMs. \trdd{Details of the sigmoid function approximation are outlined in Section \ref{LP_FA}}.\\

\noindent{\it B. Discriminative Deep Belief Networks.}~Deep Belief Networks (DBNs) are probabilistic generative models which learn to extract a deep hierarchical representation from the training data. Hinton et al. ~\cite{hinton2006fast} shows that DBNs are stacked RBMs and can be learned in a greedy manner by sequentially learning RBMs. Using Contrastive Divergence, the first layer is trained as an RBM with the input of the DBN as its input layer and, after the first RBM is trained, the weights $\bm W^1$ are fixed, as well as representations of $\bm h^1$. The binary states of the first hidden \sd{unit layer} are then used as inputs training the second RBM for $\bm W^2$. \fdic{Iterating this way layer-by-layer, the DBN can be trained by greedily training RBMs.}

\begin{figure}
\centerline{\includegraphics[width=18pc]{./FIGS/DDBN-Architecture.png}}
\caption{\footnotesize \textsf{\textbf{Discriminative DBN Architecture.} The DDBN architecture consists of stacked RBMs with a DRBM at the classification layer. Similar to the DRBM, the visible neurons $\bm v$ are split into input neurons $\bm x$ and class neurons $\bm c$. Each hidden layer $\bm h^l$, where $l \in \{ 1, \cdots, L\}$, is greedily trained layer by layer~\cite{hinton2006fast}.}}
\label{DDBN-arch}
\end{figure}

Similar to the DRBM, a DBN can \trd{also} be used as a discriminative model by \tb{concatenating a classification layer to the final hidden layer}, as shown in Fig.~\ref{DDBN-arch}.
\tb{Discriminative DBNs (DDBNs) can \trd{be used to} perform classification using either Free Energy or Gibbs Sampling, as described \trdd{for DRBMs}.}
\tb{With the Free Energy method, we compute the activation probability of all hidden units across layers. Following this, the classification result is given by the class $\bm c$ that has the minimum Free Energy across all classes based on the activation values from the last hidden layer $L$.}
\tb{Using Gibbs Sampling, the binary activation state of each class given a visible vector $\bm x$ is found by repeated sampling of all hidden neurons across all layers and class neurons. The correct classification result is given by the class with the highest activation frequency.}
Similar to the DRBM,  
our hardware implementations of DDBNs use Gibbs Sampling \trdd{for on-chip classification.} \\ 

\noindent{\it C. Network Architecture.}
\label{sec:framework}
We use the Discriminative DBN to perform classification on the MNIST dataset, which contains 60k training samples and 10k test samples of 28x28 gray-scale handwritten digits.
We binarize the images using a fixed threshold of 0.5. 
 Throughout this paper, our DRBM and DDBN naming \trdd{conventions} \trdd{denote} the amount of neurons in each hidden layer bottom up, e.g. DDBN-100-200 denotes a 2 layer DDBN with 100 neurons in the first hidden layer and 200 neurons in the second hidden layer. Our analysis in this paper is based on the following architectures:
 
\begin{itemize}
\item 300 neuron budget: DRBM-300, DDBN-100-200
\item 600 neuron budget: DRBM-600, DDBN-100-200-300
\end{itemize}
 
 
 
 

\section{Limited Precision \& Function Approximation}
\label{LP_FA}
\sd{Discriminative DBNs implemented with full precision weights and biases and nonlinear sigmoidal activation functions cannot be directly implemented for inference on finite precision digital hardware platforms.}
\icc{We therefore study the effect of fixed-point representations and approximate sigmoid functions on DDBN classification performance.} \trd{These approximations form the basis of our AX-DBN framework used for the implementation of DDBNs in finite precision embedded hardware.} \\

\noindent{\it A. Limited Precision Approximation.} \label{sec:VP}
Fixed-point is preferred to floating-point representation because the latter requires extensive area and power for
digital hardware implementations~\cite{ly2009high}. As shown in Fig. \ref{S1toS3}, there are three steps in the computation where fixed-point variables can be used to map the network onto its limited precision (LP) version:
\begin{itemize}
\item {\rm LP1} limits the precision of weight $\bm{W}$ and bias $\bm{b}$
\item {\rm LP2} limits the precision of the quantity $\bm{Wx + b}$
\item {\rm LP3} limits the precision of $\sigma (\bm{Wx + b})$ 
\end{itemize}


\begin{figure}[b]
\centerline{\includegraphics[width=11pc]{./FIGS/LP.png}}
\caption{\footnotesize \textsf{\textbf{Stages of Limited Precision Approximation.} We can exploit limited precision (LP) approximations at various stages in the network. 
}}
\label{S1toS3}
\end{figure}

\noindent{\it B. Sigmoid Function Approximation.} Approximate implementations and comparisons of 
sigmoid activation functions designed for digital hardware exist in the literature~\cite{tommiska2003efficient,amin1997piecewise}. Here we use PLAN, a piecewise linear approximation of the sigmoid function proposed by Amin et al.~\cite{amin1997piecewise} in our design

{\footnotesize\begin{equation}
y =  \left\{ 
\begin{aligned}
& f(x) && x>0 \\
& 1-f(|x|) && x\leq 0
\label{eq:PLAN}
\end{aligned}
\right.
\end{equation}}
\normalsize

\vspace*{-0.15cm}
where 

\vspace*{-0.2cm}
{\footnotesize
$$
f(x) =  \left\{ 
\begin{aligned}
& 1 && x>5 \\
& 0.03125x+0.84375 &&2.375<x\leq 5 \\
& 0.125x+0.625 && 1<x\leq 2.375 \\
& 0.25x + 0.5 && 0 \leq x \leq 1
\end{aligned}
\right.
$$
}
\normalsize

\noindent{Only addition and shift operations are involved in the approximation given by Eq.~\ref{eq:PLAN} which allows for a relatively inexpensive implementation in digital hardware with minimal accuracy loss.} 

\section{Neuron Ordering using Criticality Analysis}
\label{sec: crit analysis}

A neuron is said to be \textit{critical} if a network's performance is significantly degraded by random noise injected on said neuron, otherwise it is said to be \textit{resilient}~\cite{zhang2015approxann}. The performance of a neural network is invariant to lowering the precision of, or even removing, resilient neurons, whereas it is significantly degraded otherwise. 
In the AxNN~\cite{venkataramani2014axnn} and ApproxANN~\cite{zhang2015approxann} design methodologies, which use deterministic feedforward networks, the Euclidean distance between the classification output and true label is used as the loss function for both training and criticality analysis.
The magnitude of the average loss over all training samples is used to characterize the criticality of each neuron.
 
 
\sdd{For the case of stochastic neural networks, such as DRBMs and DDBNs, in this paper we perform the following steps for criticality analysis:}
\begin{itemize}
\item \sdd{After performing stochastic learning, we estimate the criticality of individual neurons using a gradient-based backpropagation approach as outlined in~\cite{venkataramani2014axnn} and~\cite{zhang2015approxann}. 
However, unlike deterministic models, DRBMs and DDBNs are fundamentally stochastic models which can in principle be used for both discrimination and generation tasks. 
Keeping this perspective of a fully probabilistic framework, we propose the use of Cross Entropy as a loss function for determining critical neurons using gradient-based backpropagation.}
\item \sdd{Inference in DRBMs and DDBNs is performed by Gibbs Sampling as described in Section \ref{sec: dbn theory} for ease of implementation in digital hardware. However, for criticality analysis we represent the binary states of the neurons in these networks with their activation probabilities, which are continuous between 0 and 1. }
\end{itemize}

\vspace{-0.2cm}
\icc{In the Cross Entropy (CE) loss function, $y_j$ and $a_j^c$ denote the one-hot ground truth and softmax class prediction probabilities, respectively.} 

\vspace*{-0.7cm}
{\small
\begin{center}
$$
\mathbb{L}_{\rm CE} = -\sum_{j} y_j \ln\left(a^c_j\right)
$$
\end{center}
}

\vspace{-0.2cm}
\fdic{The derivative of a given loss function with respect to the value of a hidden neuron relates that neuron's contribution to classification error caused by bitwidth and functional approximations at that neuron.
Using CE, the error sensitivity of loss due to corruption of neuron $j$ in layer $\bm h^L$ (denoted as $h^L_j$) of a DDBN with $L$ layers is defined in Eq.~\ref{eq: crit deriv}, where $a^c_i$ is defined by the softmax output in Eq.~\ref{eq:DRBMPc} and $z^L_i = b^c_i + \sum_j h^L_j W^c_{ij}$.} 

{\footnotesize 
\begin{equation}
\label{eq: crit deriv}
\begin{aligned}
\frac{\partial \mathbb{L}_{\rm CE}}{\partial h^L_j}
&= \sum_i \frac{\partial \mathbb{L}}{\partial a^c_i} \frac{\partial a^c_i}{\partial z_i^L} \frac{\partial z_i^L}{\partial h^L_j}\\
&= \sum \limits _{i} (-\frac{y_i}{a^c_i}) \cdot \left( \frac{\partial \text{softmax} (z_i^c)}{\partial z_i^c}  \frac{\partial \sum \limits_{k} (h_{k}^{L}W_{ik}^c + b_l^c)}{\partial h_j^L}\right) \\
&= \sum \limits_{i} (-\frac{y_i}{a^c_i}) \cdot a^c_i (\mathbbm{1}_{i=j}-a^c_j ) W^c_{ij} 
\end{aligned}
\end{equation}}

\vspace{-0.3cm}
\fdic{The error sensitivity of neuron $j$ in hidden layers $\bm{h}^l$, where $l < L$,  is computed by Eq.~\ref{eq: hidden neurons}. The binary states of individual neurons $h^l_i$ are approximated by their sigmoid output $a^l_i$. Here, $\sigma(z^l_i)=1/(1+\exp(-z^l_i))$}.

{\footnotesize
\begin{equation}
\begin{aligned}
\frac{\partial \mathbb{L_{\rm CE}}}{\partial h^l_j}
&= \sum_{i} \frac{\partial \mathbb{L}}{\partial a^{l+1}_i} \frac{\partial a^{l+1}_i}{\partial h^l_j}\\
&= \sum \limits_{i} \frac{\partial \mathbb{L}}{\partial a^{l+1}_i} \cdot \sigma(z^{l+1}_j) ( 1 - \sigma(z^{l+1}_j) ) W^{l+1}_{ij} 
\end{aligned}
\label{eq: hidden neurons}
\end{equation}}

\vspace{-0.3cm}
For $k$ samples of training data, the criticality score for neuron $j$ in hidden layer $\bm{h}^l$ with $N_{\bm{h}^l}$ neurons, where $l \in \{1,\cdots,L \}$, is given by:

\vspace{-0.2cm}
{\small
\begin{equation}
\label{eq: gradient based score - normed}
    \text{criticality}(h^l_j) = \frac{\E[{h^l_j}]}{\frac{1}{N_{\bm{h}^l}} \sum_{i=1}^{N_{\mathbf{h}^l}} \E[{h^l_j}]} 
\end{equation}
}

\noindent{where {\small $\E[{h^l_j}] =   \frac{1}{k} \sum_{\mathbf{x, y} \, \in \, \text{data}} \left\vert \frac{\partial \mathbb{L_{\rm\tiny{CE}}}}{\partial h^l_j} \right\vert$}.}

The hidden neurons are then ranked in order from least to most critical by the magnitude of these obtained scores. \par

\begin{figure*}[t]
\centerline{\includegraphics[width=22pc]{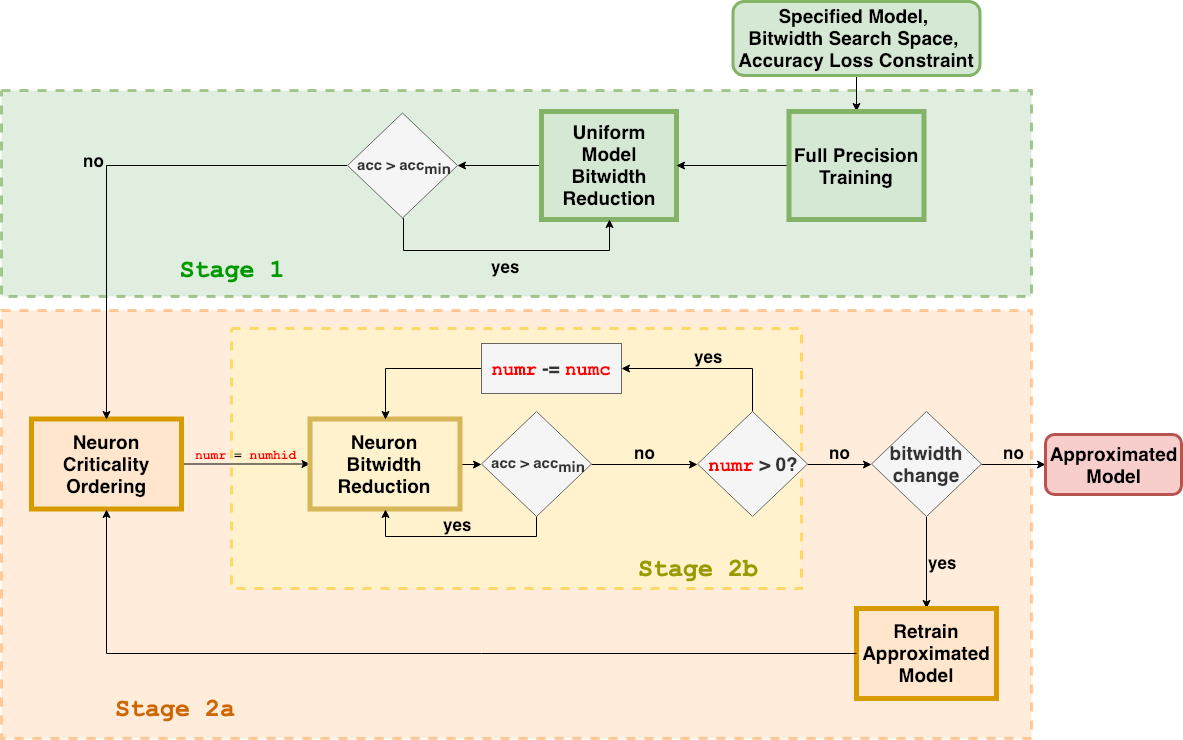}}
\caption{\footnotesize{\textsf{\textbf{Cloud Server Network Approximation Design Flow}. \icc{The approximation algorithm can be broken into two stages: (1) uniform bidwidth reduction and (2) neuron criticality analysis with systematic bitwidth reduction and retraining.} The algorithm takes in a \trdd{specified} model, a set bitwidth search space, and an accuracy loss constraint as inputs then returns an approximated energy-efficient model. All algorithm hyperparameters are shown in red. The number of hidden neurons is given by \textbf{\rm \bf numhid}, the number of resilient neurons to approximate is given by \textbf{\rm \bf numr}, and the search step size if given by \textbf{\rm \bf numc}.}}} 
\label{approxdbn flow chart}
\end{figure*}

\section{AX-DBN Design Metholodogy}
The design methodology illustrated by Fig.~\ref{fig: macro info flow graph} is \trdd{composed} of two parts: (1) a cloud-based \trd{training and} approximation process that optimizes the precision of individual neurons for inference on hardware; (2) inference performed locally on embedded hardware.\\

\vspace*{-0.15cm}
\noindent{\it A. Cloud-based Model \trd{Training and} Approximation.}~
The approximation algorithm is a scalable framework that can be applied to any fully connected network of variable width and depth. 
Figure~\ref{approxdbn flow chart} illustrates the approximation flow chart, starting with a user-specified accuracy loss constraint \trdd{with respect to a full precision implementation} and \trdd{a} \trd{set of allowed bitwidths for a specified DRBM/DDBN model} then ending with its approximated counterpart.
\fdic{Algorithm~\ref{alg: design} formally describes this approximation procedure and can be summarized by two stages:}
\begin{enumerate}
    \item[1)] Full precision training and \trd{uniform} bitwidth reduction
    \item[2)] Neuron bitwidth reduction using (a) neuron criticality analysis and retraining and (b) limited precision neuron approximation
\end{enumerate} 

\trd{Stage 1 first trains the specified model at full precision then subsequently reduces bitwidth uniformly across all weights and biases until it can no longer maintain the accuracy loss constraint}.
Stage 2 first analyzes and rank orders hidden neurons according to a given criticality metric\trdd{,} then individually approximates \sd{these} neurons \sd{based on the criticality ranking} and, finally retrains the limited precision model to allow for further network approximation. In our experiments, hidden neurons can be represented at 64-bit $Q8.56$, 16-bit $Q8.8$, 12-bit $Q6.6$, 8-bit $Q4.4$, or 4-bit $Q1.3$ precision.\footnote{We found that these $Qm.n$ bitwidths gave the best accuracy when approximating the models uniformly.} 
Here, we use $Qm.n$ to denote the fixed-point precision format with $m$ integral bits  (including sign bit), and $n$ fractional bits.  Additionally, the algorithm can prune the neuron completely.
\fdic{Class neurons are represented at 16-bit $Q8.8$ precision and not considered in our optimization due to their relatively minor impact on power savings when compared to that arising from approximating the larger number of hidden neurons considered in our architectures.}\\

\begin{algorithm}[t]
\renewcommand{\algorithmicrequire}{\textbf{Input:}}
\renewcommand{\algorithmicensure}{\textbf{Output:}}
\caption{\footnotesize \textsf{\textbf{AX-DBN Approximation Algorithm.} {\small The approximation algorithm greedily explores neuron bitwidth distributions given a specified model ($\text{\rm model}_0$), an accuracy constraint ($\text{\rm acc}_{\rm min}$), and a set bitwidth search space ({\rm bw}). Algorithm hyper-parameters include the desired neuron step size ({\rm numc}), the number of approximated resilient neurons ({\rm numr}), the total number of hidden neurons ({\rm numhid}). The summed difference in all neurons bitwidths between successive criticality-retraining iterations is $\Delta \text{\rm sumbit}$.}}}
\begin{algorithmic}[1]
\REQUIRE \textit{specified model} [$\text{model}_0$], \textit{min accuracy} [$\text{acc}_{\rm min}$], \textit{bitwidth search space} [bw]
\STATE $\text{model} \gets fullprecisionTraining(\text{model}_0)$
\STATE $\text{model}_{\rm ax} \gets uniformApprox(\text{model})$
\WHILE{$\Delta \text{sumbit} > 0$}
\STATE $\text{order} \gets criticalityScore(\text{model}_{\rm ax})$
\STATE $\text{numr} \gets \text{numhid}$
\WHILE{$\text{numr}>0$ and $\text{acc} > \text{acc}_{\rm min}$}
\STATE $\text{model}_{\rm ax} \gets neuronApprox(\text{model}_{\rm ax}, \text{numr}, \text{order})$
\STATE $\text{numr} \gets \text{numr} - \text{numc}$
\ENDWHILE
\STATE $\text{model}_{\rm ax} \gets retrain(\text{model}_{\rm ax})$
\ENDWHILE
\ENSURE \textit{approximated model} [$\text{model}_{\rm ax}$]
\end{algorithmic}
\label{alg: design}
\end{algorithm}

\vspace*{-0.2cm}
\noindent{\it B. Inference in Hardware.}~
\trd{After the cloud-based approximation procedure is completed, we download the limited precision weights and biases onto embedded hardware and \trdd{perform} Gibbs Sampling for local inference}, as described in Section~\ref{sec: dbn theory}. \ic{As depicted in Fig.~\ref{DDBN}, our hardware implementations use a pipelined architecture consisting of $L+1$ stages with one stage for each of the $L$ hidden layers and a classification layer, respectively.}

\begin{figure}[t]
\centerline{\includegraphics[width=14pc]{./FIGS/HardwareInferencePath.png}}
\caption{\footnotesize \textsf{\textbf{\footnotesize DDBN Inference on Local Embedded Hardware.} In our hardware implementations, we use Gibbs Sampling to propagate binary activation states from the input layer $\bm x$ to the classification layer $\bm c$ and use the PLAN approximation for each sigmoid output. \tb{Each hidden layer performs random sampling using a pseudo-random pattern generator (PRPG) \trdd{implemented in hardware}.}}}
\label{DDBN}
\end{figure}

\section{Power Model for Computation Workload and Memory Access}


\icc{To perform inference on a DDBN with $x$ visible units, $c$ class units, $k$ data samples, and $L$ hidden layers each with $N_{h_l}$ hidden neurons, a chip needs to read network parameters and input data from memory and to write classification results to memory}. Each weight or bias is represented by an $q$-bit fixed-point number, each input sample is a $x$-dim binary vector, and each classification result is represented by a $c$-dim binary vector. {In this model, $q$ refers to $m+n$ in a $Qm.n$ representation and \fdic{the number of neurons at $q$-bit precision in hidden layer $\bm h^l$ is given by $N^q_{h_l}$}}.



The power \trd{for} off-chip to on-chip data transfer $DT$ is given by Eq.~\ref{eq: dt power model}, where the power consumption for reading and writing 1 bit of data is given by $A$.
The computation workload $CW$ is defined by Eq.~\ref{eq: cw power model} in terms of multiply-accumulate operations, where power consumption for a $q$-bit ALU operation is given by $B_q$. 
$A$ is obtained by modeling in CACTI~\cite{hlabs}. We calculate $B_q$ as the average accumulator power consumption \trd{using} \sd{Verilog implementations of our specified DRBM and DDBN architectures} at $q$-bit precision \sd{with Synopsys EDA tools using a 65 nm standard cell library}. We estimate the total power consumed by an AX-DBN approximated model to be the combination of its data transfer and computation workload, i.e $DT$ + $CW$. In our implementation, $q$ takes \trdd{the following} values: $\{4,8,12,16,64\}.$ 



\section{Experimental Results}

An effective \trd{criticality} metric should be able to accurately determine hidden neurons that can be approximated using lower precision representations while maintaining \sd{specified} classification accuracy above a desired threshold. \fdic{Our simulations show that the criticality measure based on Cross Entropy used in this paper achieves this objective.} \trd{Owing} to the stochasticity of our models, \trd{we run 200 Monte Carlo iterations} to \trd{obtain distributions of} power savings and average neuron bitwidth.  \trd{A different random seed is used in each Monte Carlo iteration} to initialize the weights and biases of a new model and train it using Contrastive Divergence (CD) on the MNIST dataset. 
We then compare the performance of criticality metric driven realizations versus that of random ordering.\par 

We explore the effectiveness of our AX-DBN framework across the \sd{DRBM and} DDBN architectures presented in Section~\ref{sec:framework}. For \trd{an even} comparison, we \trd{fix} the \trd{total} number of neurons as we increase the depth of the network - i.e. DDBN-100-200 and DRBM-300 will both have a fixed budget of 300 neurons. \ic{In our experiments, we \sd{first} train each network using Contrastive Divergence on \trd{the MNIST dataset}, we then \trd{perform our criticality-driven approximations on} each network, and finally evaluate the approximated model's performance. \nic{Reducing the bitwidth of all neurons uniformly results in significant accuracy degradation, as shown in Table~\ref{tbl: uniform accuracy}. With each median and mean, we provide the Inter-Quartile Range (IQR) and standard deviation in parenthesis, respectively. Using criticality-driven neuron approximation and pruning, we are able to significantly reduce the average neuron bitwidth while maintaining user-specified accuracy loss constraints with respect to ideal full precision implementations, as shown in Tables~\ref{tbl: 1pct bit acc} and \ref{tbl: 5pct bit acc}.}}

\begin{figure*}
    \centering
    \includegraphics[width=0.3\textwidth]{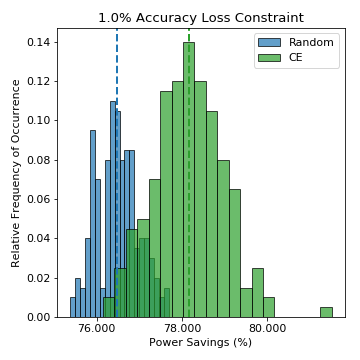}
    \includegraphics[width=0.38\textwidth]{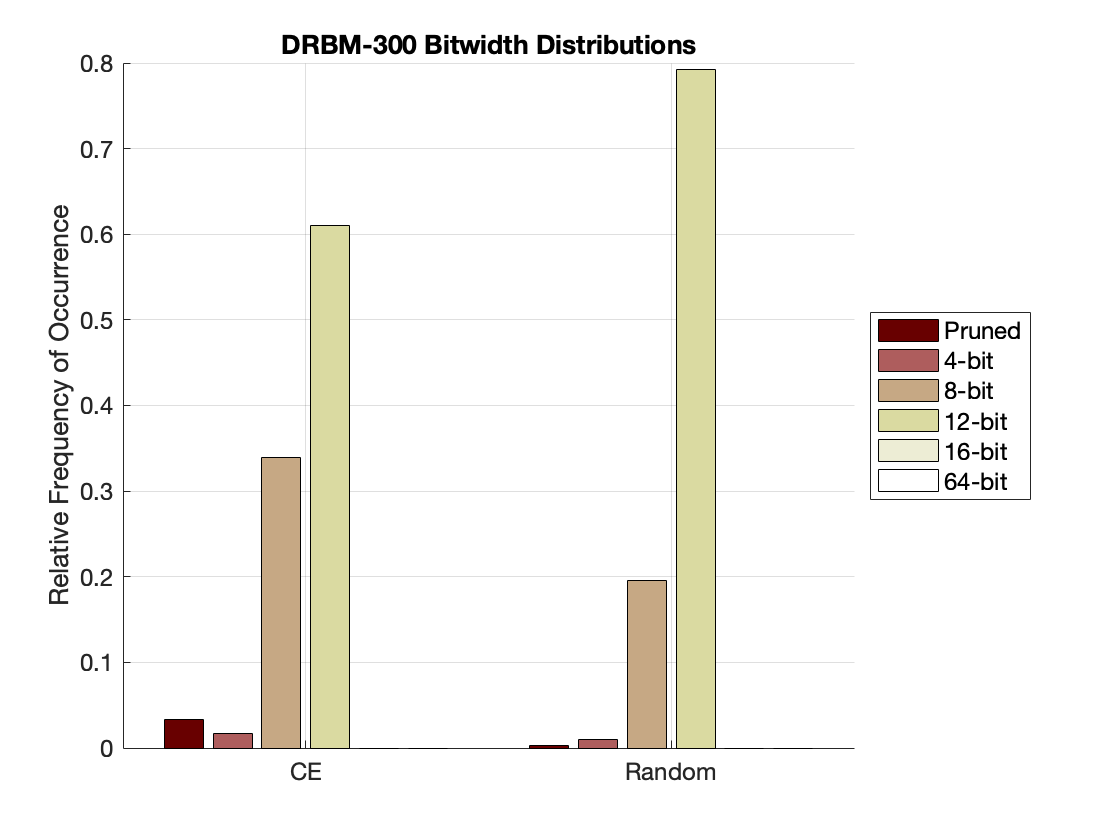}
    \includegraphics[width=0.3\textwidth]{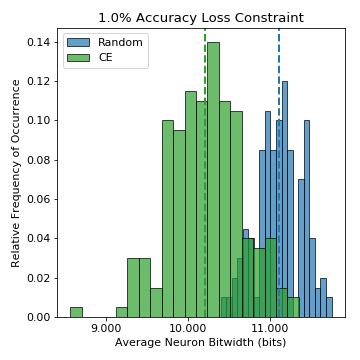} \\

    \includegraphics[width=0.3\textwidth]{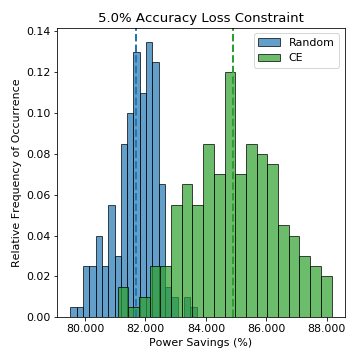}
    \includegraphics[width=0.38\textwidth]{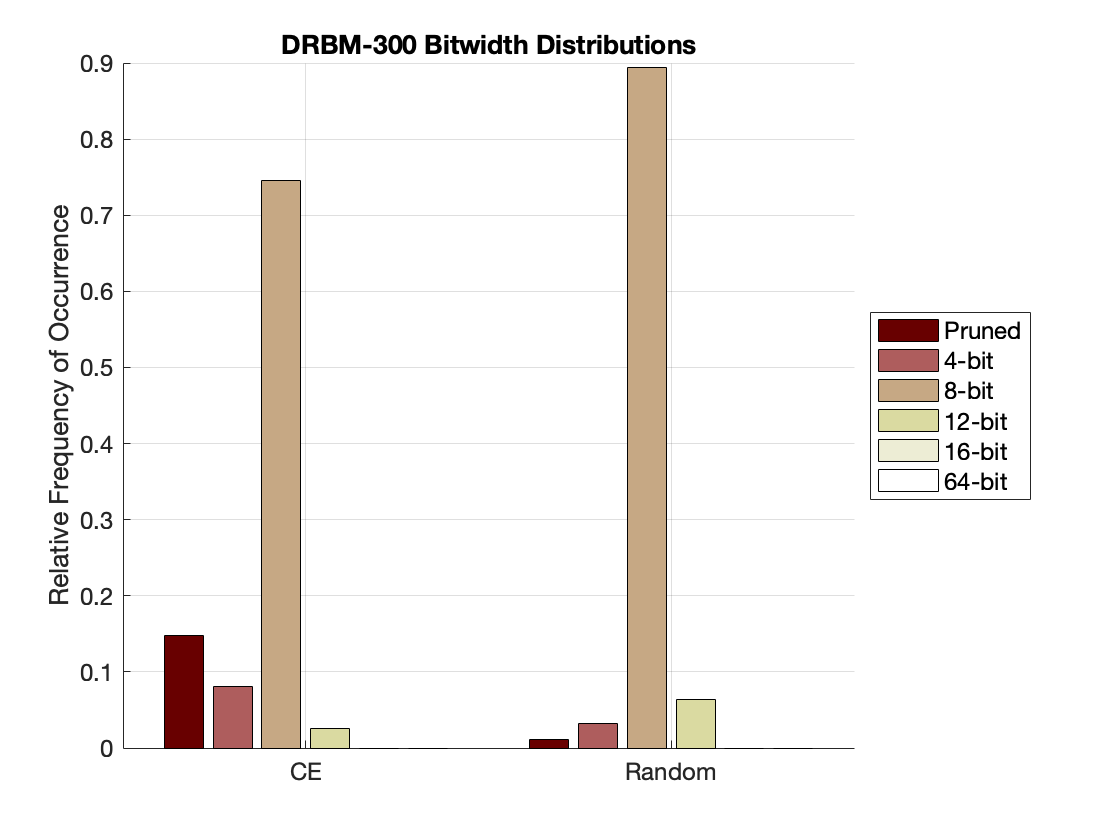}
    \includegraphics[width=0.3\textwidth]{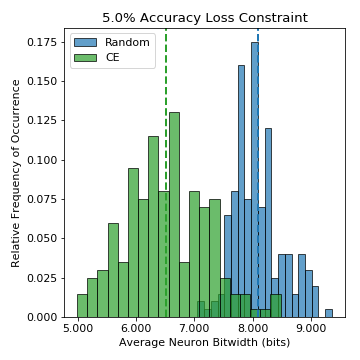}
\caption{\footnotesize \textsf{\textbf{DRBM-300.} Relative power savings (left), neuron bitwidth distribution (middle) and average neuron bitwidths (right).}}
\label{fig: rbm-300 results}%
\end{figure*}

\begin{figure*}
\centering
\includegraphics[width=0.3\textwidth]{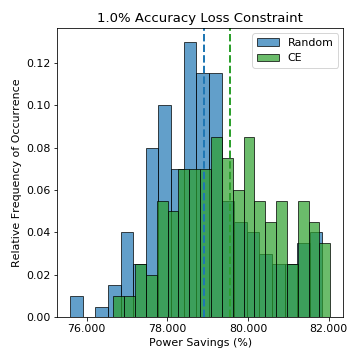}
\includegraphics[width=0.38\textwidth]{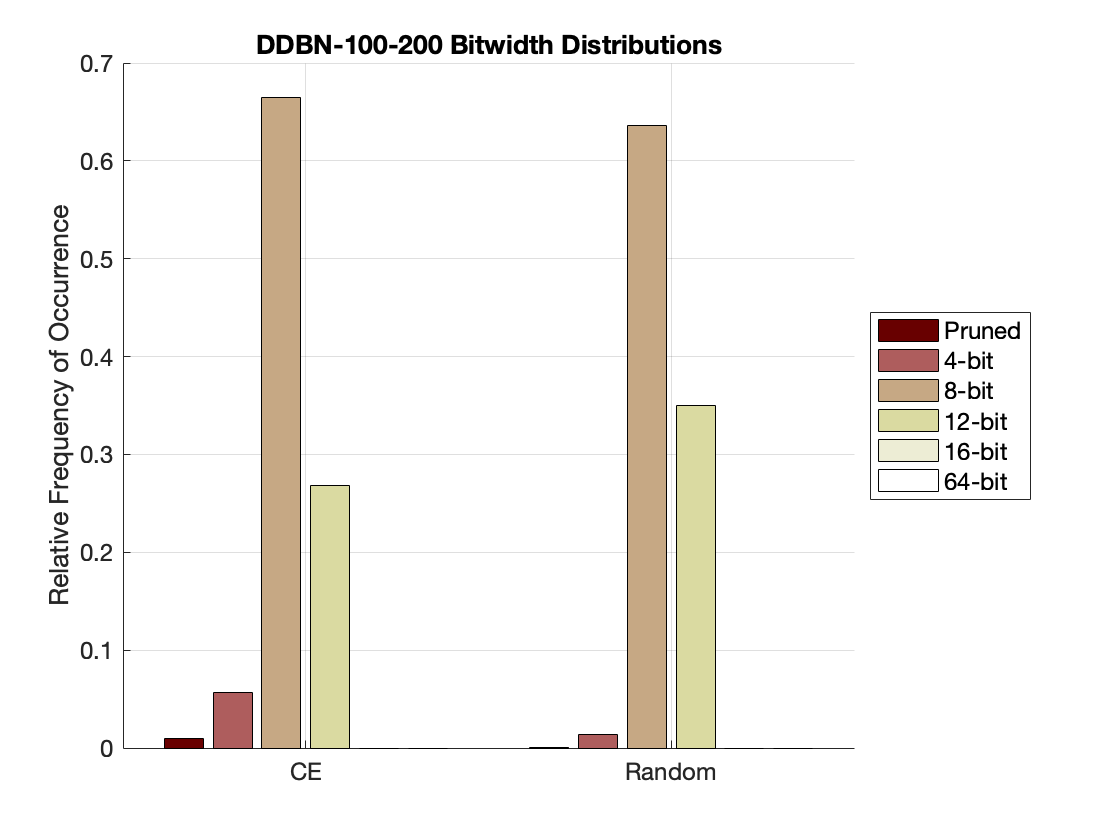}
\includegraphics[width=0.3\textwidth]{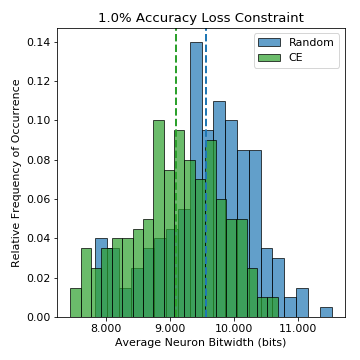}\\

\includegraphics[width=0.3\textwidth]{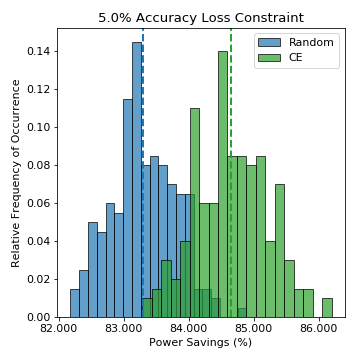}
\includegraphics[width=0.38\textwidth]{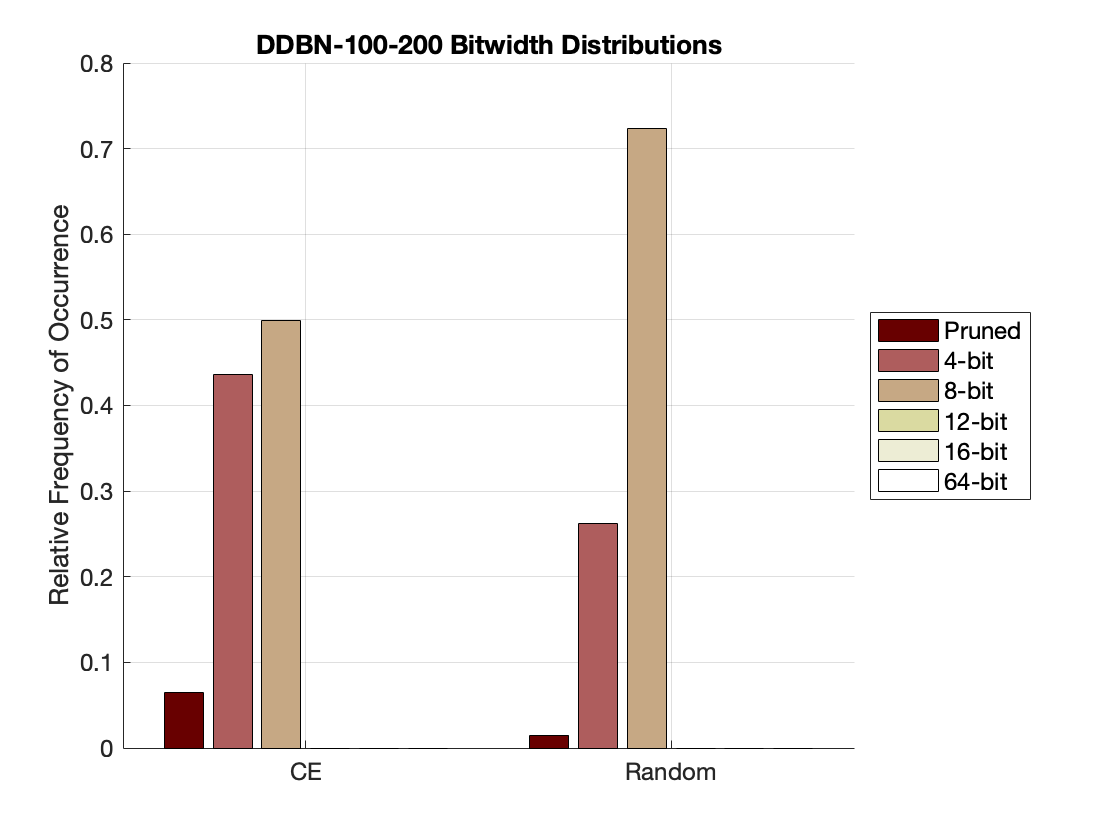}
\includegraphics[width=0.3\textwidth]{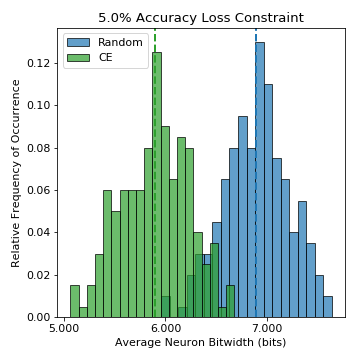}
\caption{\footnotesize \textsf{\textbf{DDBN-100-200.} Relative power savings (left), neuron bitwidth distribution (middle) and average neuron bitwidths (right).}}
\label{fig: dbn-100-200 results}%
\end{figure*}

\begin{figure*}
\centering
\includegraphics[width=0.3\textwidth]{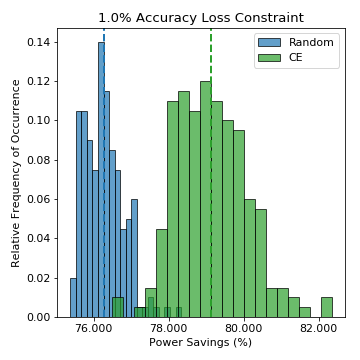}
\includegraphics[width=0.38\textwidth]{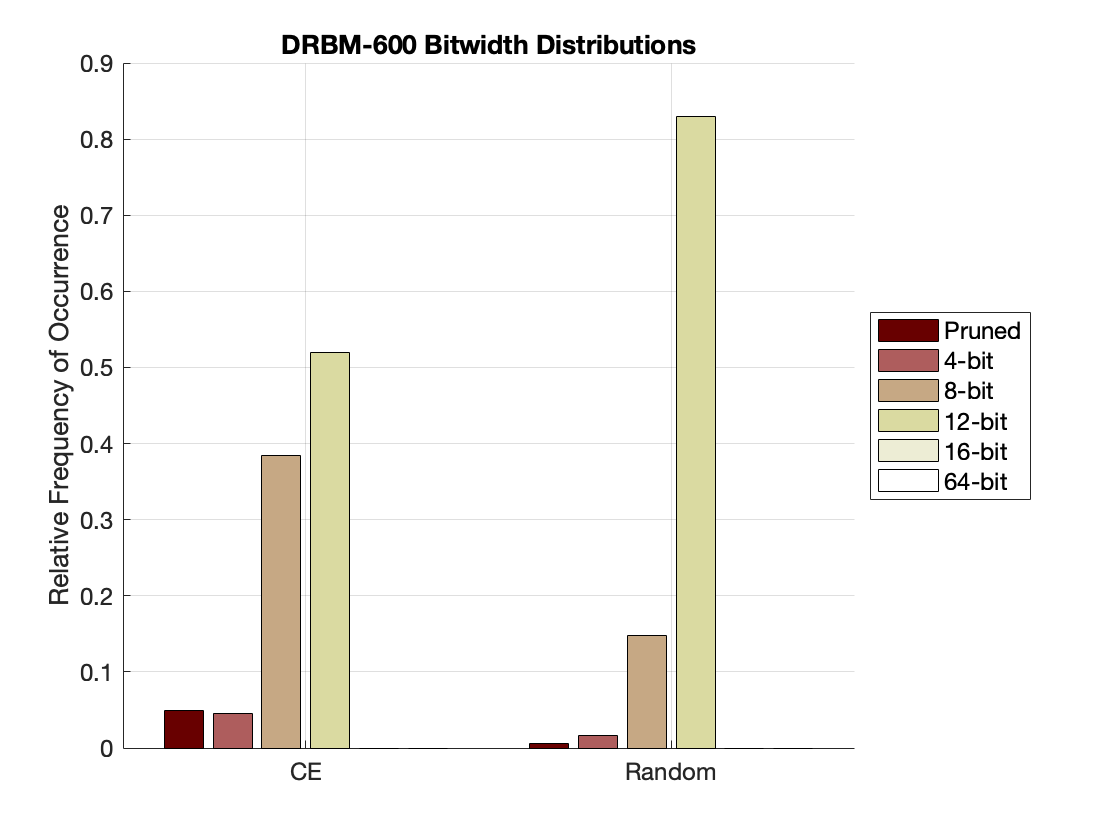}
\includegraphics[width=0.3\textwidth]{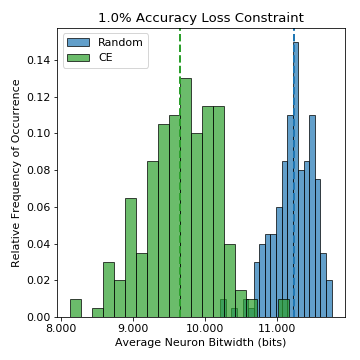}\\
\includegraphics[width=0.3\textwidth]{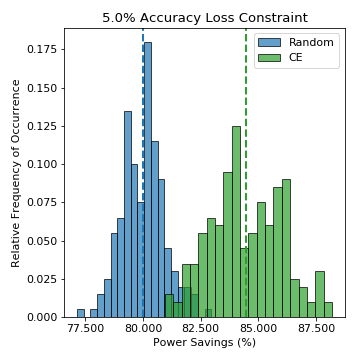}
\includegraphics[width=0.38\textwidth]{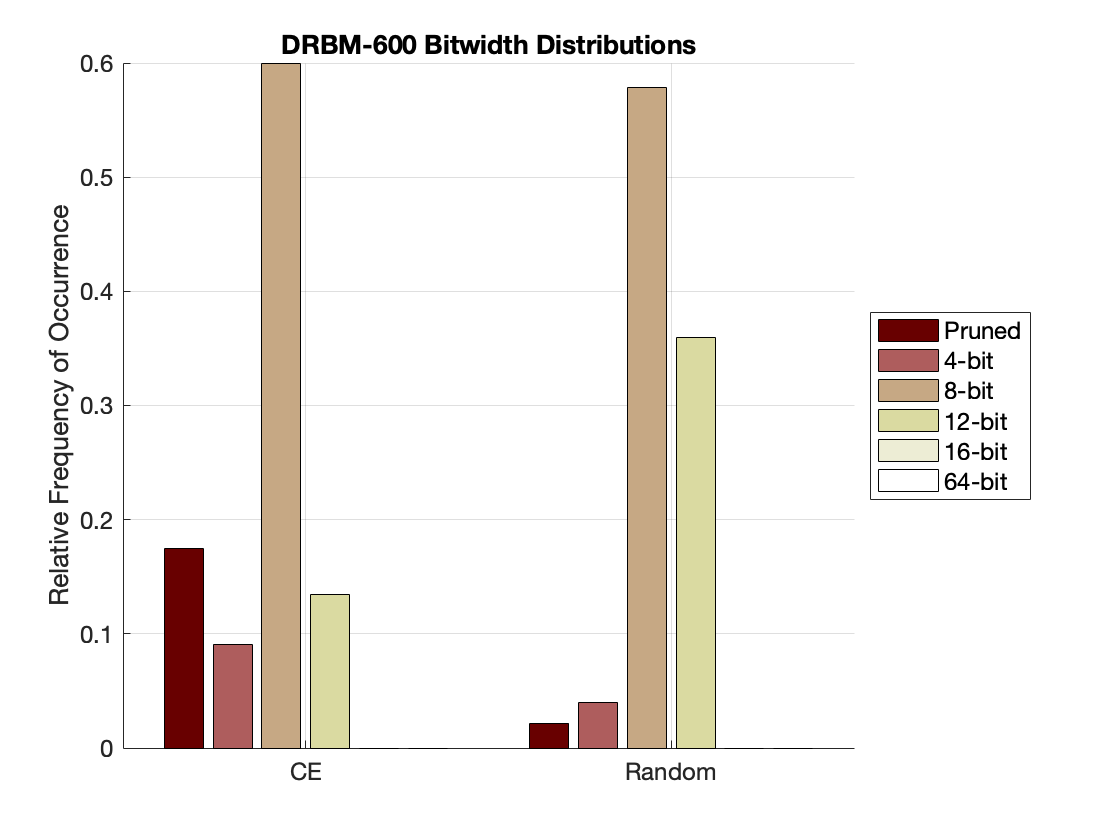}\includegraphics[width=0.3\textwidth]{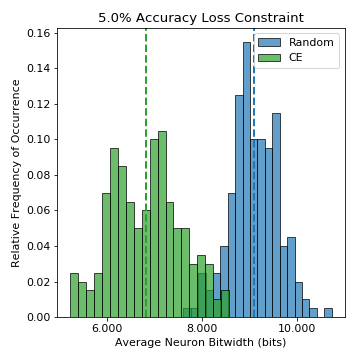}
\caption{\footnotesize \textsf{\textbf{DRBM-600.} Relative power savings (left), neuron bitwidth distribution (middle) and average neuron bitwidths (right).}}%
\label{fig: rbm-600 results}%
\end{figure*}

\begin{figure*}
\centering
\includegraphics[width=0.3\textwidth]{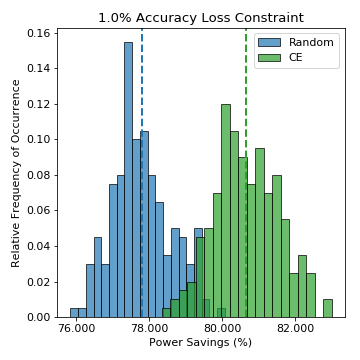}
\includegraphics[width=0.38\textwidth]{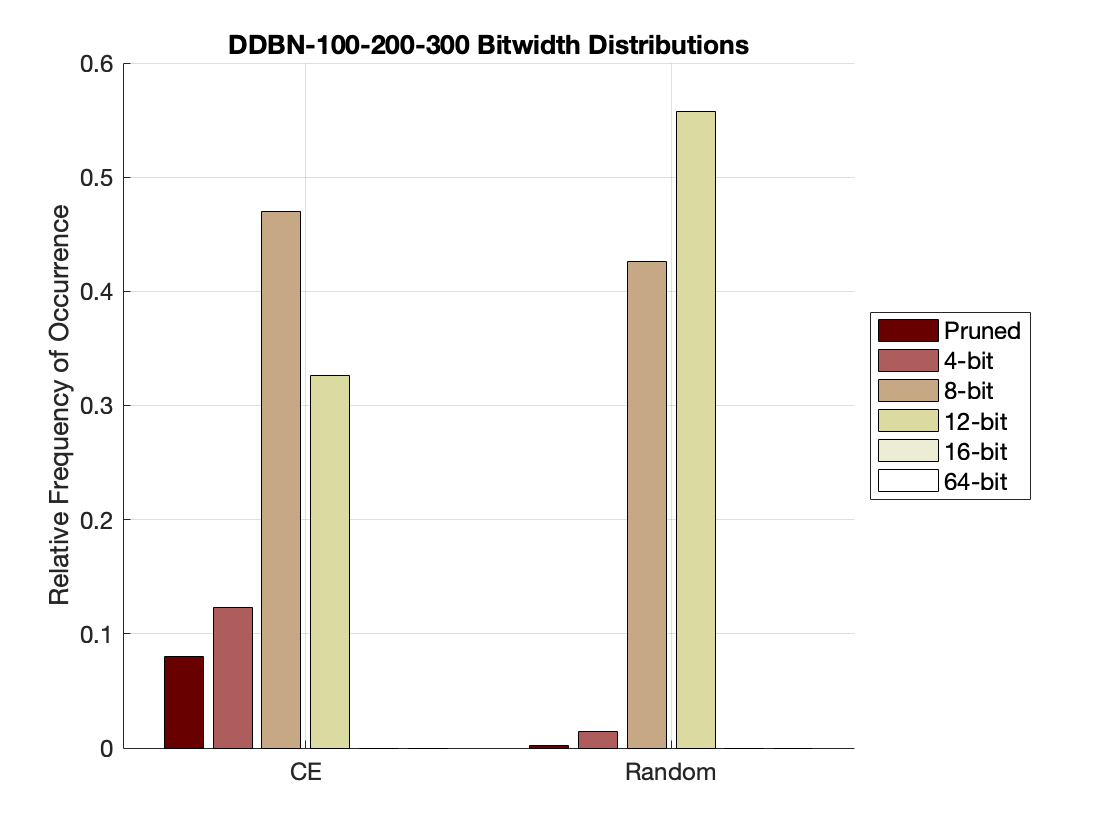}
\includegraphics[width=0.3\textwidth]{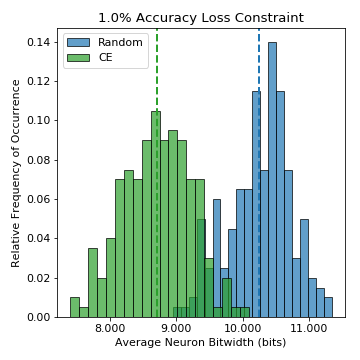}\\
\includegraphics[width=0.3\textwidth]{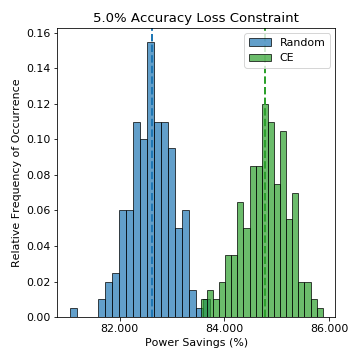}
\includegraphics[width=0.38\textwidth]{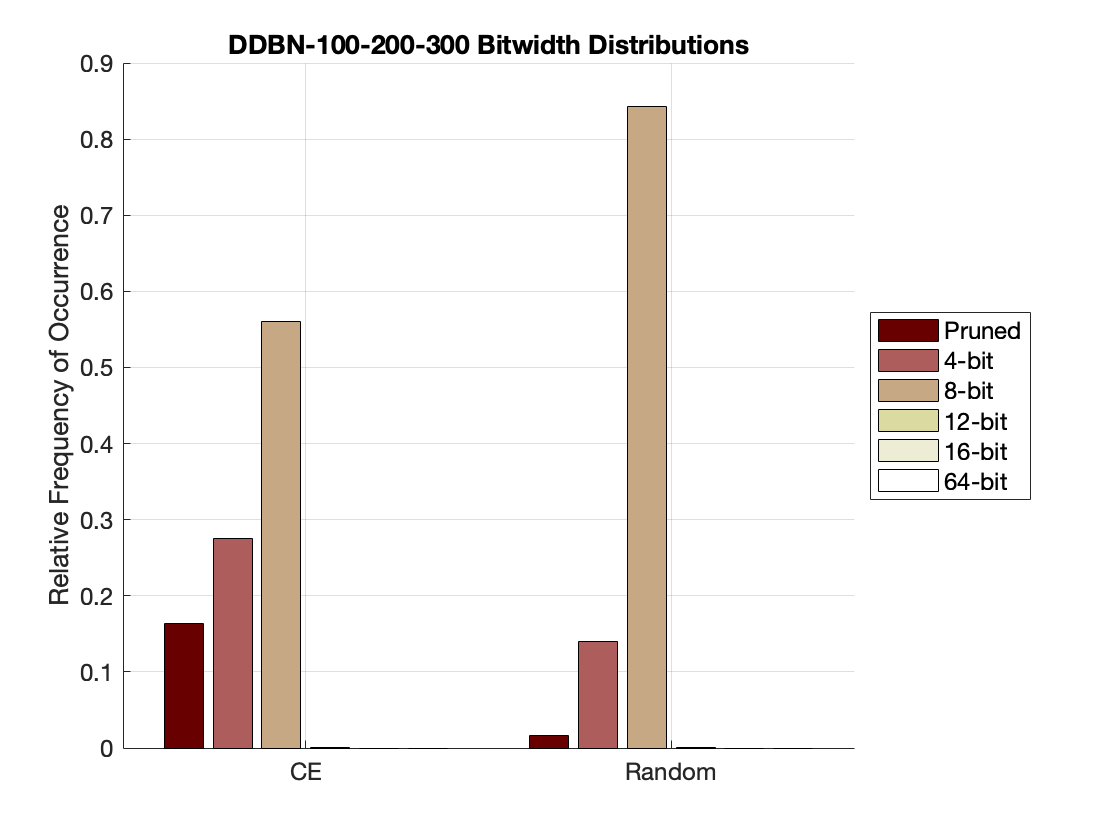}
\includegraphics[width=0.3\textwidth]{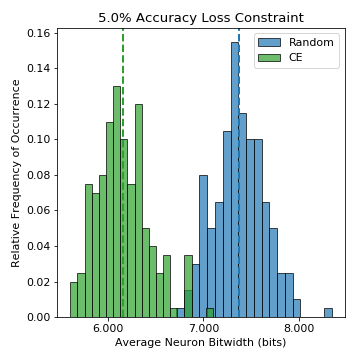}
\caption{\footnotesize \textsf{\textbf{DDBN-100-200-300.} Relative power savings (left), neuron bitwidth distribution (middle) and average neuron bitwidths (right).}}
\label{fig: dbn-100-200-300 results}%
\end{figure*}

Figures~\ref{fig: rbm-300 results}, \ref{fig: dbn-100-200 results}, \ref{fig: rbm-600 results} and \ref{fig: dbn-100-200-300 results} visualize the power savings with respect to a 64-bit implementation, the distribution of neuron bitwidths, and average neuron bitwidth for 1\% and 5\% accuracy loss constraints for different network architectures. 
We observe that approximated networks realized using Cross Entropy driven critical neuron determination yield higher average power savings and lower average neuron bitwidths when compared to those realized using random neuron ordering. 
Overall, Cross Entropy is an effective metric for hidden neuron criticality determination across accuracy loss constraints and stochastic network architectures.

{\small\vspace*{-0.7cm}
\begin{strip}
\begin{align}
\label{eq: dt power model}
    DT &= A \left( (x + 1) \sum_{q=1}^{64} q N^q_{h_1} + \sum_{l=1}^{L-1} (N_{h_l} + 1) \sum_{q=1}^{64}(q N^q_{h_{(l+1)}})+16(N_{h_L} + 1)c + k(x + c) \right)
    \\
\label{eq: cw power model}
    CW &= k \left( (x+1) \sum_{q=1}^{64}(B_q N^q_{h_1}) + \sum_{l=1}^{L-1} (N_{h_l} + 1) \sum_{q=1}^{64} B_q  N^q_{h_{(l+1)}} + B_{16} (N_{h_L} + 1) c \right)
\end{align}
\end{strip}}



\begin{table}[t]
\fontsize{6}{8}\selectfont
\centering
\begin{tabular}{|c|c|c|c|c|c|c|}
\hline
\rowcolor[HTML]{C0C0C0} 
Architecture  & \begin{tabular}[c]{@{}c@{}}4-bit \end{tabular} & \begin{tabular}[c]{@{}c@{}}8-bit \end{tabular} & \begin{tabular}[c]{@{}c@{}}12-bit \end{tabular} & \begin{tabular}[c]{@{}c@{}}16-bit \end{tabular} & \begin{tabular}[c]{@{}c@{}}64-bit \end{tabular} \\ \hline
\cellcolor[HTML]{EFEFEF}DRBM-300 & 54.5 (5.3) & 88.6 (1.1) & 94.0 (0.3) & 94.3 (0.2) & 94.3 (0.3) \\ \hline
\cellcolor[HTML]{EFEFEF}DDBN-100-200 & 73.7 (4.7) & 94.3 (0.5) & 95.9 (0.2) & 95.9 (0.2) & 96.0 (0.2) \\ \hline
\cellcolor[HTML]{EFEFEF}DRBM-600 & 54.2 (4.2) & 82.7 (3.0) & 95.0 (0.2) & 95.3 (0.2) & 95.3 (0.2) \\ \hline
\cellcolor[HTML]{EFEFEF} DDBN-100-200-300 & 58.8 (7.4) & 93.0 (0.8) & 95.6 (0.2) & 95.6 (0.2) & 95.9 (0.2) \\ \hline\hline
\cellcolor[HTML]{EFEFEF}DRBM-300 & 54.6 (3.8) & 88.5 (0.8) & 94.0 (0.2) & 94.3 (0.2) & 94.3 (0.2)\\ \hline
\cellcolor[HTML]{EFEFEF}DDBN-100-200 & 74.0 (3.8) & 94.3 (0.4) & 95.9 (0.1) & 95.9 (0.2) & 95.9 (0.1) \\ \hline
\cellcolor[HTML]{EFEFEF}DRBM-600 & 54.0 (3.0) & 82.5 (2.3) & 95.0 (0.2) & 95.3 (0.2) & 95.3 (0.2) \\ \hline
\cellcolor[HTML]{EFEFEF} DDBN-100-200-300 & 58.7 (5.3) & 92.8 (0.7) & 95.6 (0.2) & 95.6 (0.2) & 95.9 (0.2) \\ \hline
\end{tabular}
\caption{\footnotesize \textsf{\textbf{Classification Accuracy at Uniform Model Bitwidths}. We run 200 Monte Carlo iterations to measure the median (top) and mean (bottom) test accuracy of each architecture when reducing the bitwidth of each neuron uniformly. With each median and mean, we provide the IQR and standard deviation in parenthesis, respectively.}}
\label{tbl: uniform accuracy}
\end{table}

\begin{table}[t]
\fontsize{6}{8}\selectfont
\centering
\begin{tabular}{|c|c|c|c|c|c|c|}
\hline
\rowcolor[HTML]{C0C0C0} 
Architecture  & \begin{tabular}[c]{@{}c@{}} FP \end{tabular} & \begin{tabular}[c]{@{}c@{}} Random \end{tabular} & \begin{tabular}[c]{@{}c@{}} CE \end{tabular} \\ \hline
\cellcolor[HTML]{EFEFEF}DRBM-300 & 94.3 (0.3) & 93.7 (0.7) $\vert$ 11.1-bit (0.4) & 93.6 (0.5) $\vert$ 10.2-bit (0.6) \\ \hline
\cellcolor[HTML]{EFEFEF}DDBN-100-200 & 96.0 (0.2) & 95.0 (0.2) $\vert$ 9.63-bit (0.9) & 94.8 (0.2) $\vert$ 8.72-bit (0.9) \\ \hline
\cellcolor[HTML]{EFEFEF}DRBM-600 & 95.3 (0.2) & 94.7 (0.8) $\vert$ 11.3-bit (0.3) & 94.5 (0.6) $\vert$ 9.68-bit (0.7) \\ \hline
\cellcolor[HTML]{EFEFEF} DDBN-100-200-300 & 95.9 (0.2) & 94.8 (0.4) $\vert$ 10.3-bit (0.6) & 94.8 (0.3) $\vert$ 8.72-bit (0.7) \\ \hline\hline
\cellcolor[HTML]{EFEFEF}DRBM-300 & 94.3 (0.2) & 93.8 (0.5) $\vert$ 11.1-bit (0.3) & 93.7 (0.4) $\vert$ 10.2-bit (0.4) \\ \hline
\cellcolor[HTML]{EFEFEF}DDBN-100-200 & 95.9 (0.1) & 95.0 (0.2) $\vert$ 9.57-bit (0.7) & 94.8 (0.2) $\vert$ 8.71-bit (0.7) \\ \hline
\cellcolor[HTML]{EFEFEF}DRBM-600 & 95.3 (0.2) & 94.9 (0.5) $\vert$ 11.2-bit (0.3) & 94.6 (0.4) $\vert$ 9.66-bit (0.5) \\ \hline
\cellcolor[HTML]{EFEFEF} DDBN-100-200-300 & 95.9 (0.2) & 94.8 (0.2) $\vert$ 10.3-bit (0.5) & 94.8 (0.2) $\vert$ 8.71-bit (0.5) \\ \hline
\end{tabular}
\caption{\footnotesize \textsf{\textbf{Classification Accuracy after AX-DBN Criticality-Driven Network Approximation with a 1\% Accuracy Loss Constraint}. We run 200 Monte Carlo iterations to measure the median (top) and mean (bottom) test accuracy and average neuron bitwidth of each architecture when approximating with AX-DBN given a 1\% accuracy loss constraint. With each median and mean, we provide the IQR and standard deviation in parenthesis, respectively. The classification accuracy of the full precision (FP) model on the test dataset is given by the first column. FP denotes the full precision 64-bit floating point model.}}
\label{tbl: 1pct bit acc}
\end{table}


\begin{table}[!h]
\fontsize{6}{8}\selectfont
\centering
\begin{tabular}{|c|c|c|c|c|c|c|}
\hline
\rowcolor[HTML]{C0C0C0} 
Architecture  & \begin{tabular}[c]{@{}c@{}} FP \end{tabular} & \begin{tabular}[c]{@{}c@{}} Random \end{tabular} & \begin{tabular}[c]{@{}c@{}} CE \end{tabular}  \\ \hline
\cellcolor[HTML]{EFEFEF}DRBM-300 & 94.3 (0.3) & 91.8 (1.5) $\vert$ 8.00-bit (0.6) & 92.4 (1.0) $\vert$ 6.46-bit (1.0) \\ \hline
\cellcolor[HTML]{EFEFEF}DDBN-100-200 & 96.0 (0.2) & 92.4 (1.0) $\vert$ 6.91-bit (0.4) & 92.5 (0.8) $\vert$ 5.92-bit (0.5) \\ \hline
\cellcolor[HTML]{EFEFEF}DRBM-600 & 95.3 (0.2) & 93.5 (1.1) $\vert$ 9.05-bit (0.6) & 94.4 (0.7) $\vert$ 6.86-bit (1.1) \\ \hline
\cellcolor[HTML]{EFEFEF} DDBN-100-200-300 & 95.6 (0.2) & 92.4 (0.9) $\vert$ 7.36-bit (0.4) & 92.3 (0.8) $\vert$ 6.13-bit (0.4) \\ \hline\hline
\cellcolor[HTML]{EFEFEF}DRBM-300 & 94.3 (0.2) & 91.7 (0.9) $\vert$ 8.09-bit (0.4) & 92.2 (0.8) $\vert$ 6.52-bit (0.7) \\ \hline
\cellcolor[HTML]{EFEFEF}DDBN-100-200 & 95.9 (0.1) & 92.4 (0.6) $\vert$ 6.89-bit (0.3) & 92.4 (0.6)  $\vert$ 5.89-bit (0.3) \\ \hline
\cellcolor[HTML]{EFEFEF}DRBM-600 & 95.3 (0.2) & 93.7 (0.7) $\vert$ 9.10-bit (0.5) & 94.3 (0.6) $\vert$ 6.81-bit (0.7) \\ \hline
\cellcolor[HTML]{EFEFEF} DDBN-100-200-300 & 95.9 (0.2) & 92.3 (0.7) $\vert$ 7.38-bit (0.3) & 92.2 (0.6) $\vert$ 6.15-bit (0.3) \\ \hline
\end{tabular}
\caption{\footnotesize \textsf{\textbf{Classification Accuracy after AX-DBN Criticality-Driven Network Approximation with a 5\% Accuracy Loss Constraint}. We run 200 Monte Carlo iterations to measure the median (top) and mean (bottom) test accuracy and average neuron bitwidth of each architecture when approximating with AX-DBN given a 5\% accuracy loss constraint. With each median and mean, we provide the IQR and standard deviation in parenthesis, respectively. The classification accuracy of the full precision (FP) model on the test dataset is given by the first column. FP denotes the full precision 64-bit floating point model.}}
\label{tbl: 5pct bit acc}
\end{table}

\section{CONCLUSIONS}
In this paper, we propose a systematic approximation methodology for stochastic Discriminative Restricted Boltzmann Machines (DRBMs) and Discriminative Deep Belief Networks (DDBNs) to optimize power consumption subject to the constraint of maintaining user-specified classification accuracy. This work extends criticality analysis from the domain of deterministic neural networks to the realm of stochastic networks. Our procedure involves two key steps: (1) The use of criticality analysis to rank order neurons based on their contribution to network performance; (2) The use of greedy retraining to optimize neuron bitwidths under accuracy constraints. \fdic{Our results show that Cross Entropy can be used as an effective metric for determining neuron criticality in stochastic network approximation and yields lower average neuron bitwidth representations as well as higher savings in power consumption when compared to random criticality ordering of neurons.} Our future research will extend our methodology to optimize the power consumption of hardware implementations of generative neural networks for purposes such as image generation, denoising, and infilling.\par


\bigskip
\centerline{A\footnotesize{CKNOWLEDGEMENTS}}

\medskip
This work was supported in part by NSF awards CNS-1730158, ACI-1540112,  ACI-1541349, OAC-1826967, the University of California Office of the President, and the California Institute for Telecommunications and Information Technology's Qualcomm Institute (Calit2-QI). Thanks to CENIC for the 100Gpbs networks. We would also like to thank Professor Andrew Kahng, Jonas Chan and Chih-Yin Kan at UC San Diego for providing assistance with Verilog implementations and power measurements.



%


\end{document}